\documentclass{article}

\usepackage{amsmath}
\usepackage{amssymb}
\usepackage{graphicx}
\usepackage{pstricks}
\usepackage{epic,eepic}
\usepackage{fancyhdr}
\usepackage{verbatim}
\usepackage{latexsym}
\usepackage{enumerate}
\usepackage{mathrsfs}

\newcommand{\PP}{{\mathcal{P}}}
\newcommand{\GG}{{\mathcal{G}}}

\newcommand{\NN}{{\mathbf{N}}}

\newcommand{\FF}{{\mathcal F}}
\newcommand{\BB}{{\mathcal B}}

\newtheorem{prop}{Proposition}
\newtheorem{thm}[prop]{Theorem}

\newtheorem{lem}[prop]{Lemma}
\newtheorem{dfn}[prop]{Definition}

\begin{document}

\title{Nonlocal, noncommutative diagrammatics and the linked cluster Theorems}
\author{Christian Brouder
\footnote{Institut de Min\'eralogie et de Physique des Milieux Condens\'es,
CNRS UMR 7590, Universit\'es Paris 6 et 7, IPGP, 140 rue de Lourmel,
75015 Paris, France.}, Fr\'ed\'eric Patras \footnote{Laboratoire J.-A. Dieudonn\'e, CNRS UMR 6621,
Universit\'e de Nice,
Parc Valrose, 06108 Nice Cedex 02, France.}}

\date{\today}
\maketitle
\begin{abstract}
Recent developments in quantum chemistry, perturbative quantum field theory, statistical physics or stochastic differential equations require the introduction of new families of Feynman-type diagrams. These new families arise in various ways. In some generalizations of the classical diagrams, the notion of Feynman propagator is extended to generalized propagators connecting more than two vertices of the graphs. In some others (introduced in the present article), the diagrams, associated to noncommuting product of operators inherit from the noncommutativity of the products extra graphical properties.

The purpose of the present article is to introduce a general way of dealing with such diagrams. 
We prove in particular a ``universal'' linked cluster theorem and introduce, in the process, a Feynman-type ``diagrammatics'' that allows to handle simultaneously nonlocal (Coulomb-type) interactions, the generalized diagrams arising from the study of interacting systems (such as the ones where the ground state is not the vacuum but e.g. a vacuum perturbed by a magnetic or electric field, by impurities...) or
Wightman fields (that is, expectation values of
products of interacting fields).
Our diagrammatics seems to be the first attempt to encode in a unified algebraic framework such a wide variety of situations. 

In the process, we promote two ideas. First, Feynman-type diagrammatics belong mathematically to the theory of linear forms on combinatorial Hopf algebras. Second, linked cluster-type theorems rely ultimately on M\"obius inversion on the partition lattice. The two theories should therefore be introduced and presented accordingly.

Among others, our theorems encompass the usual versions of the theorem (although very different in nature, from Goldstone diagrams in solid state physics to Feynman diagrams in QFT or probabilistic Wick theorems). 
\end{abstract}
%
%\tableofcontents

\section{Introduction}

The attempts to clarify the mathematical framework underlying quantum chemistry, solid state physics, quantum field theories (QFT) and connected topics such as  -for example, the grounds for the functional approaches, the principles underlying renormalization- as well as attempts to deepen our current understanding of widely used techniques (effective Hamiltonians, adiabatic limits...) are often hindered by very basic questions regarding the underlying mathematical methods. This is particularly obvious when it comes to developing new tools, see e.g. our studies of enhanced algorithms and formulas for the computation of eigenstates of Hamiltonians when the (unperturbed) ground state is degenerate 
\cite{BP-09,BMP-10,BDPZ10}, or of the combinatorics of one-particle-irreducibility for interacting systems \cite{BP09}.

The present article focusses on diagrammatics. We argue that Feynman-type diagrammatics belong mathematically to the theory of linear forms on combinatorial Hopf algebras, which allows to generalize the theory to a much wider setting than the classical one.
We cover the usual examples corresponding to vacuum expectations over commutative products of fields where the propagators are represented by edges -this includes for example the various Goldstone-type diagrams and the perturbative expansions parametrized by Feynman diagrams in quantum field theories. However we go beyond and cover the case of expectations over general states, which requires the introduction of generalized diagrams \cite{Djah,BP09}. The same combinatorics happens to provide a pictorial description of cumulants in probability. The Hopf algebraic approach also allows to study expectations of products of free fields (Wightman fields) and derive new Feynman diagram expansions in this setting.

In a second part, we prove general linked cluster theorems, covering all
these cases and making as explicit as possible the link between the
elementary algebra underlying the theorems and the graphical content
(that relies on connectedness in the graph-theoretical sense). Notice
that we avoid deliberately functional methods (see e.g.\cite{Itzykson}):
although very efficient to derive the classical QFT linked cluster
theorem (using a generating function in terms of an external source),
they lack the generality and simplicity of the combinatorial proof.
Moreover, they cannot deal with noncommutative Feynman diagrams
because functional derivatives commute.
In the process, we promote another idea.  Namely, linked cluster-type theorems rely ultimately on M\"obius inversion on the partition lattice.

\section*{Acknowledgements} We are particularly grateful to R. Stora. Many letters and documents (among which \cite{Stora08}) he sent us were the initial incentive for the present article -that was conceived to bridge the computations in \cite{BP09} with other approaches to quantum field computations and find some suitable mathematical framework for the (much more advanced) problems these documents suggest. In particular, we aimed at developping a mathematical framework to deal with products of Wightman fields and their average values over general states, one of the problems this article addresses.

\section*{Notation} We use the following convention: since various products will be defined along the article (such as $\ast$ or $\odot$), when we want to emphasize what product is used in an exponential, a logarithm or any other operation, we put the product symbol in exponent, so that $\log^\ast$ emans that we compute the logarithm using the $\ast$ product, $x^{\odot n}$ means that we compute the $n$-th power of $x$ using the $\odot$-product, and so on.

\section{Free combinatorial Hopf algebras}
The objects we will be interested in are combinatorial Hopf algebras in the sense of Joni and Rota \cite{Joni}, that is, bialgebras which coproduct is of combinatorial nature (obtained by ``splitting'' generating symbols according to combinatorial rules encoded by remarkable ``section coefficients'' --in high-energy physics, these coefficients correspond roughly to the symmetry factors of Feynman graphs). More specifically, we will be interested in families of Hopf algebras corresponding physically to bosonic or fermionic systems, to the usual algebraic structure of quantum fields (equipped with a commutative product such as the normal or time-ordered product) and to Wightman fields. 

Since our results are more general than what would be required by applications to quantum systems, we state them in full generality and will show later how they specialize to particular physical systems or mathematical problems.
Let $X$ be a fixed ordered set, $X=\{x_1,...,x_n,...\}$. In most applications, $X$ will be infinite and countable, so that the reader may think to $X$ as the set of natural numbers. 

The notion of combinatorial Hopf algebra, goes back to \cite{Joni}. The general notion is ill-defined in the litterature (there are many natural candidates, but at the moment no convincing general definition). We choose here a simple and relatively straightforward definition suited for our purposes that reflects some of the natural properties one expects from the notion when the underlying algebra is free commutative or free associative.

\begin{dfn}
We call free commutative (resp. free) \it combinatorial Hopf algebra \rm any connected graded commutative (resp. associative) and cocommutative Hopf algebra $H$ such that:
\begin{itemize}
\item (Freeness) As an algebra, $H$ is the algebra of polynomials (resp. of tensors) over a doubly indexed (finite or countable) set of formal, commuting (resp. noncommuting), variables $\phi_{i}(x_S)$ where $S$ runs over finite subsets of $X$ and $i=1...n_k$, where $k=|S|$ and where the sequence of the $n_k$, $i\in \NN$ is a fixed sequence of integers.
%$$\phi_1(x_1),...,\phi_1(x_n),...;...;\phi_k(x_1),...,\phi_k(x_n),...$$
\item (Equivariance) The structure maps are equivariant with respect to maps induced by substitutions in $X$. In other terms, any substitution (that is, any set automorphism) $\sigma$ induces a Hopf algebra automorphism of $H$ which action on the generators is defined by $\sigma(\phi_{i}(x_S)):=\phi_{i}(x_{\sigma(S)})$.
\end{itemize}
\end{dfn}

The $\phi_{i}(x_S)$ are called the \it generators \rm of $H$, the (commutative or associative) monomials in the $\phi_{i}(x_S)$ form a linear basis of $H$ and are therefore called the \it basis elements\rm . 

We do not look for the outmost generality, and many of our constructions and definitions can be extended in a fairly straightforward way to more general systems. For example, one might consider free partially commutative Hopf algebras which generators $\phi_{i}(x_S)$ would satisfy partial commutation rules (e.g. all the $\phi_j(x_i)$ would commute for a fixed $i$, but without $\phi_j(x_i)$ and $\phi_k(x_l)$ commuting for $i\not= l$, and so on). The so-constructed Hopf algebras can make sense in various applications and inherit all the properties of free commutative or free combinatorial Hopf algebras that are required for the forthcoming reasonings, we refer e.g. to \cite{PR2002} for details on the subject.

Some further remarks are in order.
Recall first that, by the Leray theorem (see e.g. Prop 4 in \cite{Patras93}), any connected graded commutative Hopf algebra $H$ is free commutative, so that the assumption that $H$ is freely generated as a commutative algebra comes for free when one assumes that $H$ is connected graded commutative.

The key point that makes the Hopf algebra combinatorial (and, as we shall see, suited to Feynman-type graphical reasonings) is that we assume that a set of polynomial generators is fixed and behaves nicely with respect to substitutions in $X$. 

The particular case where $n_k=0$ for $k>1$ corresponds to the classical situation where the only propagators showing up in diagrams are the ones that describe the free propagation of a particle.

Let us mention at last that a more pedantical definition of combinatorial Hopf algebras could be given in terms of vector species, following the approach to combinatorial Hopf algebraic structures in \cite{patreu04,PatrasSchocker,AguiarM}.

A very important consequence of the equivariance condition is the following.
\begin{lem} For any subset $S$ of $X$, let us write $H_S$ for the subalgebra of $H$ generated by the $\phi_i(x_T),\ T\subset S$. Then, $H_S$ is a Hopf subalgebra of $H$. 
Moreover, any automorphism $\sigma$  of $X$ induces an isomorphism of Hopf algebras from $H_S$ to $H_T$, where $T:=\sigma(S)$.
\end{lem}

The second property is a direct consequence of the first one since $\sigma$ (by definition of the induced map) induces an isomorphism of algebras from $H_S$ to $H_T$.
The proof of the first assertion follows immediately from the equivariance condition. Indeed, notice that an element of $H$ or of $H\otimes H$ is invariant by any substitution that acts as the identity on a finite subset $S$ of $X$ if and only if it is a polynomial in the $\phi_i(x_T),\ T\subset S$ (or, in $H\otimes H$, a sum of tensor products of such polynomials). Now, since,
for $T\subset S$, $\phi_i(x_T)$ is invariant by any substitution that acts as the identity on $S$, the same property holds also true of $\Delta(\phi_i(x_T))$, from which the Lemma follows.

\section{Some remarkable Hopf algebras}
Our favorite examples in view of applications to quantum chemistry, QFT
and solid state physics are simple ones. However the generality chosen
(allowing for example $n_2\not= 0$) is natural to handle nonlocal
interaction terms in the Lagrangians (think for example of the quantum
chemistry approach with Coulomb interaction). Our general approach also paves the way to a unification of QFT techniques (Feynman-type diagrammatics), umbral calculus (duality and linear forms on polynomial algebras) and combinatorics of (possibly ordered) set partitions. 

For notational simplicity, we treat only commutative algebras in the usual sense, that is we do not treat explicitly the fermionic case (Grassmann or exterior algebras/ Fermi statistics). 
However, as it is well known, there are no difficulties in switching from a bosonic to a fermionic framework -it just requires adding the right signs in the formulas, see \cite{Cassam,BrouderQG}, but handling simultaneously the commutative and anticommutative case would have required introducing consistently signs in all our formulas. For notational simplicity, we decided to stick to the bosonic, commutative, case, and to let the interested reader adapt our results to the anticommutative setting.
We use the langage of particle physics and quantum chemistry (so that the bosonic algebra coincides with the algebra of polynomials).

\begin{dfn} [Bosonic algebra]
We write ${\mathcal B}_k^X$ for the algebra of polynomials  over the set of (formal, commuting) variables $$\phi_1(x_1),...,\phi_1(x_n),...;...;\phi_k(x_1),...,\phi_k(x_n),...$$ 
where $x_i\in X$.
These algebras are naturally equipped with a coproduct $\Delta: \BB\longmapsto\BB\otimes \BB$ that makes them Hopf algebras (that is, $\Delta$ is a map of algebras). The coproduct is defined on the generators $\phi_l(x_i)$ by requiring them to be primitive, that is: $$\Delta (\phi_l(x_i)):=\phi_l(x_i)\otimes 1+1\otimes \phi_l(x_i)$$ and extended multiplicatively to $\PP$ ($\Delta(xy)=\Delta(x)\Delta(y)$).
\end{dfn}

The lower index $k$ should be thought of as the number of quantum fields showing up (in a very broad sense), whereas the $x_i$ should be thought of as points, momenta or more generally dummy integration variables. For later use, we allow $k=\infty$ (to encode the countable set of eigenstates of a given many-body Hamiltonian) and will write simply $\cal B$ for ${\cal B}_k^X$ when no confusion can arise.

\begin{dfn} [Coulomb algebra]
We write ${\mathcal P}_k^X$ for the algebra of polynomials  over the set of (formal, commuting) variables $\phi_l(x_i),\ l\leq k, \ \phi(x_{i,j})$, where $i\not= j\in X$.
The coproduct is defined by requiring the generators to be primitive.
\end{dfn}
This algebra is used in non-relativistic many-body physics
and quantum chemistry. The Coulomb interaction desribes
the force between a charge at point $x_i$ and
a charge at point $x_j$. These two points are linked by the
interaction, and a specific variable $\phi(x_{i,j})$ is
used to describe this connection.

\begin{dfn} [Tensor algebra]
We write ${\FF}_k^X$ for the algebra of noncommutative polynomials  over the set of variables $$\psi_1(x_1),...,\psi_1(x_n),...;...;\psi_k(x_1),...,\psi_k(x_n),...$$
$x_i\in X$.
The coproduct is defined by requiring the generators to be primitive.
\end{dfn}

Various other free combinatorial Hopf algebras possibly noncommutative but fitting in the general framework of the present article are described (or follow from the results) in \cite{PatrasSchocker,PatrasSchocker2}. Let us quote the Malvenuto-Reuntenauer Hopf algebra and various Hopf algebras of tree-like structures. Although the decorated version of the Connes-Kreimer Hopf algebra of trees in \cite{PatrasSchocker} may have a use for QFT, a more promissing path in that direction is certainly provided by \cite{Gurau}, the belief of which we do share : ``ultimately we think that combinatorics is the right approach to QFT and
that a QFT should be thought of as the generating functional of a certain
weighted species in the sense of \cite{Leroux}''. We restrict however the examples in the present article to well-established domains of QFT and leave further extensions to future work.

\section{Graphication}
Let us start by recalling the construction underlying diagrammatic expansions and show how it can be extended naturally in a noncommutative/nonlocal setting. We call this process ``graphication'', by analogy with the ``arborification'' process underlying tree expansions in analysis and dynamical systems \cite{Ecalle92,Menous07}. In all the article, as a tribute to the physical motivations the ground field is the field of complex numbers (although the results hold over an arbitrary field of characteristic zero).

The reduced coproduct $\overline\Delta : H\longmapsto H\otimes H$ is defined by $\forall x\in H, \overline\Delta(x):=\Delta(x)-x\otimes 1-1\otimes x$. The coproduct and reduced coproduct are coassociative in the sense that $(\Delta\otimes H)\circ \Delta=(H\otimes \Delta)\circ \Delta$ (and similarly for $\overline\Delta$). The iterated coproduct and reduced coproduct maps from $H$ to $H^{\otimes i}$ are therefore well-defined and written $\Delta^{[i]}$, resp. $\overline\Delta^{[i]}$. 

We assumed in the definition of combinatorial Hopf algebras that the coproduct is cocommutative: with Sweedler's notation $\Delta (x)=x^{(1)}\otimes x^{(2)}$, this means that $x^{(1)}\otimes x^{(2)}=x^{(2)}\otimes x^{(1)}$. Many of our forthcoming results could be adapted to the noncocommutative setting. However, this hypothesis is particularly usefulf when it comes to graphical encodings.

\begin{dfn}
The graphication map ${\mathcal G}$ is the map from $H$ to $\bigoplus\limits_n(H^{\otimes n})^{sym}\subset \bigoplus\limits_n(H^{\otimes n})$ defined by:
$${\mathcal G}:=\sum\limits_n\frac{\overline\Delta^{[n]}}{n!}.$$
\end{dfn}

Here, the superscript $sym$ in $\bigoplus\limits_n(H^{\otimes n})^{sym}$ means that the elements in the image of ${\mathcal G}$ are sums of symmetric tensor powers of elements of $H$. We will use the canonical isomorphism from covariants $\bigoplus\limits_n(H^{\otimes n})_{sym}:=\bigoplus\limits_nH^{\otimes n}/S_n$ (where $S_n$ stands for the symmetric group of order $n$) to invariants
$$[x_1| ...| x_n]\longmapsto\frac{1}{n!}\sum\limits_{\sigma\in S_n} x_{\sigma(1)}\otimes ...\otimes x_{\sigma(n)}$$ to represent invariants using the bar notation (that is, $y_1\otimes ...\otimes y_n$ in $H^{\otimes n}_{sym}$ is written $[y_1|...|y_n]$). Notice that, by definition (since we deal with covariants), for any permutation $\sigma$, $[y_1|...|y_n]=[y_{\sigma(1)}|...|y_{\sigma(n)}]$.

For example, in the bosonic algebra, abbreviating $\phi_1$ to $\phi$ (a notation we will use without further comments from now on):
$$\GG(\phi(x_1)\phi(x_2)...\phi(x_n))=\sum\limits_{\mathcal I}[\prod_{i\in {I_1}}\phi(x_{i})|...|\prod_{i\in {I_k}}\phi(x_{i})]$$
where ${\mathcal I}$ runs over all partitions $I_1\coprod...\coprod I_k$, $k=1...n$ of $[n]:=\{1,...,n\}$, with $\inf\{i\in I_j\}<\inf\{i\in I_{j+1}\}.$
Or
(isolating some components in the expansion):
$$\GG(\phi(x_1)^4\phi(x_2)^4)=[\phi(x_1)^4\phi(x_2)^4]+ 4[\phi(x_1)|\phi(x_1)^3\phi(x_2)^4]+...$$
$$+ 18[\phi(x_1)^2\phi(x_2)^2|\phi(x_1)^2\phi(x_2)^2]+...$$

$$\GG(\phi_1(x_1)\phi_2(x_1)\phi_3(x_2)^2)=[\phi_1(x_1)\phi_2(x_1)\phi_3(x_2)^2]+...$$
$$+[\phi_1(x_1)|\phi_2(x_1)\phi_3(x_2)^2]+2[\phi_1(x_1)\phi_3(x_2)|\phi_2(x_1)\phi_3(x_1)]+...$$

The same formulas hold in the tensor algebra and in the Coulomb algebras. Notice however that, in the tensor algebra, products are noncommutative, so that the order of the products in the monomials does matter. For example, with self-explanatory notation:
$$[\psi_1(x_1)|\psi_2(x_1)\psi_3(x_2)^2]\not=[\psi_1(x_1)|\psi_3(x_2)\psi_2(x_1)\psi_3(x_2)].$$

We call brackettings the terms such as $[\phi(x_1)^2\phi(x_2)^2|\phi(x_1)^2\phi(x_2)^2]$. The \it length \rm of a bracketting is the number of vertical bars $|$ plus 1. The \it support \rm of a bracketting $\Gamma$ is the set of the $x_i$ showing up in $\Gamma$. For example, $$sup([\phi_3(x_1)\phi_4(x_2)^2|\phi_1(x_1)^2\phi_2(x_8)^2])=\{x_1,x_2,x_8\}.$$
For each $x_i\in sup(\Gamma )$, we write $d_i$ for the total degree of the $\phi_j(x_i)$s in $\Gamma$ (so that for $\Gamma$ as above, $d_1=3,\ d_2=2, \ d_8=2$). For later use, we also introduce the product of two brackettings, which is simply the concatenation product:
$$[u_1|...|u_n]\cdot[v_1|...|v_n]:=[u_1|...|u_n|v_1|...|v_n].$$

These ideas and notation extend in a self-explanatory way to arbitrary combinatorial Hopf algebras. The only change regards the definition of the support and degree: $x_i$ is included in the support of a bracketting whenever there exists a pair $(j,S)$ such that $i\in S$ and $\phi_j(x_S)$ shows up in $\Gamma$ so that, for example, 
$$sup([\phi_3(x_1)\phi_4(x_2)^2|\phi_1(x_1)^2\phi_2(x_8)^2|\phi_5(x_{1,5,8,10})])=\{x_1,x_2,x_5,x_8,x_{10}\}.$$
Similarly, the degree $d_8$ of $x_8$ in this bracketting accounts for the $\phi_5(x_{1,5,8,10})$ term and is therefore $d_8=3$.

The coefficients in the right hand side of the equation defining the
graphication will be referred to as symmetry factors. They are closely
related to the structure coefficients for the coproduct (in the basis
provided by monomials in the chosen family of generators, e.g. the
$\phi_i(x_j)$ for the bosonic algebra -these coefficients were called
section coefficients by Joni and Rota \cite{Joni}) but encode also the
symmetries showing up in the coproducts. Whereas, for the bosonic and
Coulomb algebras, symmetry factors are defined without ambiguity, for
other algebras (the tensor algebra or general free combinatorial Hopf
algebras) a given bracketting may appear in the expansion of $\GG(M)$
for various monomials $M$ in the generators (for example, in the tensor
algebra, $[\psi(x_1)|\psi(x_2)]$ appears in the expansion of
$\GG(\psi(x_1)\psi(x_2))$ and of $\GG(\psi(x_2)\psi(x_1))$). In general,
we will therefore write $s_\Gamma^M$ for the symmetry factor of a
bracketting $\Gamma$ in the expansion of $\GG(M)$ and will write simply
$s_\Gamma$ in the particular case of the bosonic and Coulomb algebras (where a unique $M$ exists giving rise to such a factor).

\section{Graphical representation}
Perturbative expansions in particle and solid-state physics are
conveniently represented by various families of diagrams. Feynman
diagrams are the most popular ones, but there are plenty of other
families with construction rules often slightly different from the one
underlying Feynman diagrams. Just to mention one interesting feature,
Feynman diagrams are usually independent of the time-coordinate of the
vertices (this is because of the definition of the so-called Feynman
propagators), whereas other families showing up in solid-state physics take into account causality systematically to construct their diagrams. These ideas are particularly well-explained in \cite{Mattuck}, to which we refer, also for a comprehensive treatment of the zoology of diagrammatic expansions.

Defining a graphical representation associated to a free combinatorial
Hopf algebra depends highly on the structure and particular features of
the algebra. We define various such representations, by increasing order
of complexity, focussing only on the algebras (tensor, bosonic, Coulomb)
we have chosen to investigate in depth. The reader who needs to
construct other taylor-made graphical representations will be able to do
so easily using our recipes (for notational simplicity and to avoid
pointless pedantry, we omit to introduce the most general possible
definitions since the process of designing them is straightforward once the leading principles are understood on some examples).

Let us mention that, when two brackettings $U=[u_1|...|u_n]$ and $V=[v_1|...|v_n]$ have disjoint supports, their product corresponds to the disjoint union of the corresponding graphs (this is the usual product on Hopf algebras of Feynman graphs in QFT, see e.g. \cite{CKI}). 

\subsection{Commutative local case: bipartite graphs}

\begin{figure}
\begin{center}
\includegraphics[width=8.0cm]{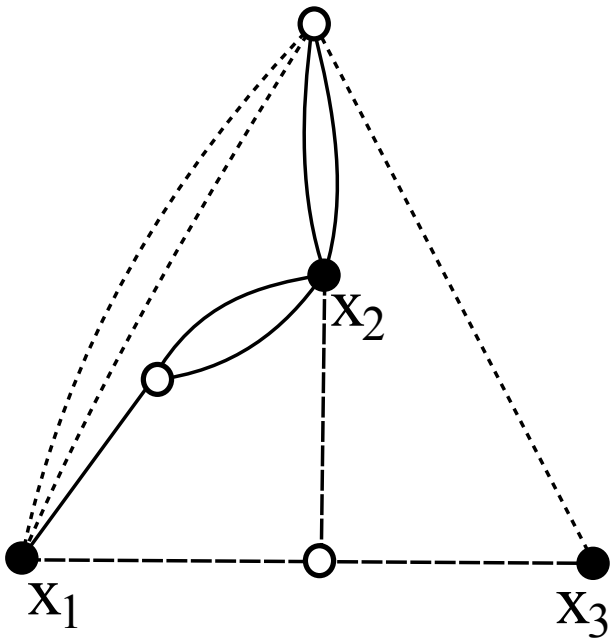}
\caption{Graph of the bracketting
$[\phi_1(x_1)\phi_1(x_2)^2|\phi_2(x_1)^2\phi_1(x_2)^2\phi_2(x_3)|
\phi_3(x_1)\phi_3(x_2)\phi_3(x_3)]$.
\label{fig1}}
\end{center}
\end{figure}

We focus in this section on the bosonic algebra.
Each Feynman bracket (say
$\Gamma=[\phi_1(x_1)\phi_1(x_2)^2|\phi_2(x_1)^2\phi_1(x_2)^2\phi_2(x_3)|
\phi_3(x_1)\phi_3(x_2)\phi_3(x_3)]$, see
fig.~\ref{fig1}) can be represented uniquely by a bipartite (non planar) graph with unoriented colored edges (i.e. by a graph with 2-coloured vertices and colored edges) according to the following rule:

\begin{enumerate}
\item For each $x_i\in sup(\Gamma )$, recall that we write $d_i$ for 
the total degree of the $\phi_j(x_i)$s in $\Gamma$. Draw a 
$x_i$-labelled black vertex with $d_i$ outgoing colored edges 
(the colors being attributed according to the indices $j$ of 
the $\phi_j(x_i)$, e.g. a 4-edge black vertex for $x_1$ with one 1-colored edge 
(solid line), two 2-colored edges (dotted line) and a 3-colored edge (dashed line).
\item Running recursively from the left to the right of $\Gamma$, 
for each term inside brackets and bars (e.g. $\phi_1(x_1)\phi_1(x_2)^2$, 
then $\phi_2(x_1)^2\phi_1(x_2)^2\phi_2(x_3)$, then $\phi_3(x_1)\phi_3(x_2)\phi_3(x_3)$), 
select randomly according to the colors 
and powers showing up in the monomials outgoing edges of the corresponding vertices 
(e.g. select one 1-colored edge from the $x_1$ vertex and two 1-colored edges 
from the $x_2$ vertex). Connect these edges to a new white vertex 
(e.g. a new white vertex with 3 outgoing colored edges).
\end{enumerate}

These edge-colored bipartite graphs will be called from now \it interaction graphs\rm . See \cite{Djah,BP09} for applications. 

Some simplifications are possible in many cases of interest. For example, when the $x_i$s are dummy integration variables and can be exchanged freely in any computation (e.g. of scattering amplitudes), the labelling of the black vertices can be omitted.

Another classical situation (encountered with scalar quantum field theories such as $\phi^3$ or $\phi^4$ and in statistical physics) is the one where $k=1$. In that case, there is just one possible color for the edges, so that the coloring of the edges can safely be omitted in the definition of the graphs. In QFT, when several fields coexist (e.g. in QED, where $k=2$), instead of using colors, 
practitioners use often different representations for the edges that depend on the fields involved (typically, in QED, edges associated to electrons are plain lines, whereas edges associated to photon propagation are represented by a succession of small waves). Of course, the use of colors or the use of different shapes for the edges are strictly equivalent and a matter of taste and habits.

Another classical simplification occurs when the only brackettings of interest for practical applications are the ones where the only monomials showing up in the brackettings are products of degree 2 of the form $\phi_i(x_j)\phi_i(x_k)$. In the corresponding graphs, the white vertices always have two outgoing edges with the same color, so that these vertices can be erased -what  remains is a graph with only black vertices and colored edges (that correspond, physically, to particle types). These are the celebrated Feynman graphs that one encounters in QFT textbooks.

From now on we will identify systematically brackettings and the corresponding interaction graphs.

\subsection{Commutative non local case: tripartite graphs}

By nonlocal case, we mean that some $n_i,\ i>1$ may be different from zero. The canonical example we have in mind is the one of QED in the solid state picture, that is with instantaneous Coulombian interactions (in the particle physics picture the interactions are local and encoded by products of fields in the Lagrangian; this corresponds to the commutative local case).

For simplicity (and in view of the most natural applications), we assume that the only brackettings of interest are those in which the monomials showing up are either monomials in the $\phi_i(x_j)$, either a $\phi_i(x_S), \ |S|>1$ (in other terms, no nontrivial products involving a $\phi_i(x_S)$ should appear in the bracketting).

Each Feynman bracket (say for example $$\Gamma=[\phi_1(x_1)\phi_1(x_2)^2|\phi_2(x_1)^2\phi_1(x_2)^2\phi_2(x_3)|\phi_3(x_{1,2,3})])$$ can be represented uniquely by a tripartite (non planar) graph with two kinds of unoriented edges according to the following rule (see Figure 2):
\begin{figure}
\begin{center}
\includegraphics[width=8.0cm]{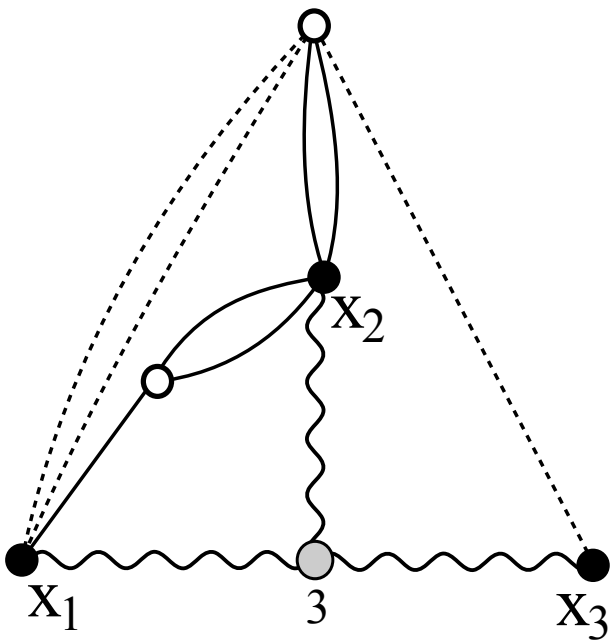}
\caption{Graph of the bracketting
$[\phi_1(x_1)\phi_1(x_2)^2|\phi_2(x_1)^2\phi_1(x_2)^2\phi_2(x_3)|
\phi_3(x_{1,2,3})]$.
\label{fig2}}
\end{center}
\end{figure}

\begin{enumerate}
\item We distinguish between the two kind of edges in the following way: the first type of edge is drawn as colored lines (the color is an index $i$ with $d_i\not=0$); the second type has no color and is drawn as a sequence of waves (as the photon propagators in QED). We call these edges respectively plain and wavy edges.
\item For each $x_i\in sup(\Gamma )$, draw a $x_i$-labelled black vertex with $d_i$ outgoing edges. Colors are attributed to the plain edges according to the indices $j$ of the $\phi_j(x_i)$, the other edges correspond to $\phi_j(x_S)$s and are wavy edges. For example, we draw a 4-edges black vertex for $x_1$ with one 1-colored edge (represented by a solid line in Fig. 2), two 2-colored edges (dotted lines) and a wavy edge. 
\item Running recursively from the left to the right of $\Gamma$, for each term inside brackets and bars (e.g. $\phi_1(x_1)\phi_1(x_2)^2$, then $\phi_2(x_1)^2\phi_1(x_2)^2$, then $\phi_3(x_{1,2,3})$), \begin{itemize}
\item If encountering a monomial involving $\phi_i(x_j)$s, proceed as in the commutative local case:
select randomly according to the colors and powers showing up in the monomials outgoing edges of the corresponding vertices (e.g. select one 1-colored edge from the $x_1$ vertex and two 1-colored edges from the $x_2$ vertex). Connect these edges to a new white vertex (e.g. a new white vertex with 3 outgoing colored edges).
\item If encountering a $\phi_j(x_S)$, select randomly a wavy edge outgoing from the $x_i, \i\in S$ black vertex, connect all these edges to new $j$-labelled grey vertex.
\end{itemize}
\end{enumerate}

These tripartite graphs will be called from now 
\emph{nonlocal interaction graphs}. The particular case of solid state physics QED enters in this framework. Moreover, the general case we consider seems new and of interest, since it should allow to treat QED computations in a complex background (e.g. with non trivial vacua). These concrete applications (that originated this work, together with questions and remarks by R. Stora) are left for future work.

Great simplifications occur in many simpler cases of interest. The first
situation encountered in pratice is the one where $n_1=n_2=1$. This
corresponds roughly to the case where one class of particles is present
(say electrons, up to a change from the bosonic to the fermionic
statistics) and where interactions are encoded by nonlocal terms
(Coulomb interaction lines). In that case, there is just one possible color for the edges, so that the coloring of the plain edges can be omitted in the definition of the graphs. Besides, since $n_2=1$, the grey vertices can also be erased, so that one ends up with bipartite graphs with two types of edges (plain and wavy). If one assumes further that the only graphs of interest are those with two outgoing edges, these edges can be also erased. One ends up with the familiar Feynman diagrams with black vertices only and two types of edges corresponding to electron propagators and Coulombian interactions.

\subsection{Noncommutative local case}
We focus in this section on the tensor algebra.
Each Feynman bracket (say $\Gamma=[\psi_1(x_2)\psi_1(x_1)\psi_1(x_2)|\psi_2(x_1)\psi_1(x_2)\psi_2(x_3)\psi_1(x_2)\psi_2(x_1)|
\psi_3(x_3)\psi_3(x_1)\psi_3(x_2)]$) can be represented uniquely by a bipartite (non planar) graph with unoriented colored and locally ordered edges (i.e. by a graph with 2-coloured vertices and colored, locally ordered edges) according to the following rule (the rule will make clear the meaning of ``local order'' that corresponds to an order on the edges reaching a white vertex)
(see fig.~\ref{fig3}):
\begin{figure}
\begin{center}
\includegraphics[width=8.0cm]{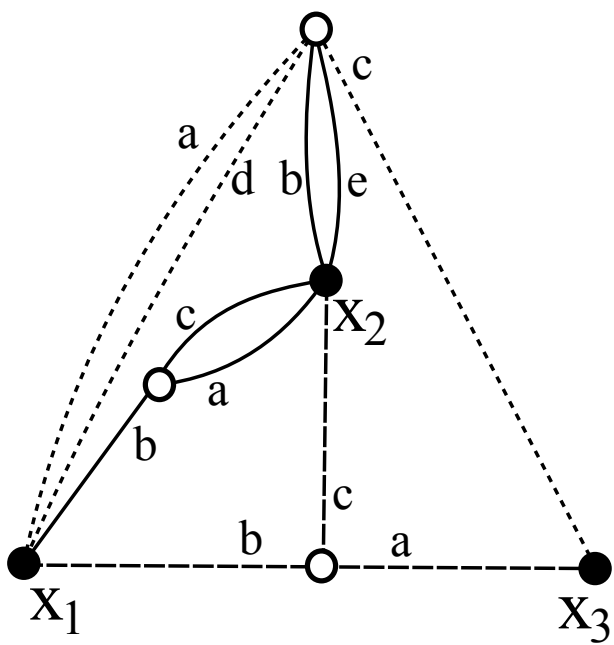}
\caption{Graph of the bracketting
$[\psi_1(x_2)\psi_1(x_1)\psi_1(x_2)|\psi_2(x_1)\psi_1(x_2)\psi_2(x_3)\psi_1(x_2)\psi_2(x_1)|\psi_3(x_3)\psi_3(x_1)\psi_3(x_2)]$.
\label{fig3}}
\end{center}
\end{figure}

\begin{enumerate}
\item Proceed first as in the commutative case: for each $x_i\in sup(\Gamma )$, recall that we write $d_i$ for the total degree of the $\phi_j(x_i)$s in $\Gamma$. Draw a $x_i$-labelled black vertex with $d_i$ outgoing colored edges (the colors being attributed according to the indices $j$ of the $\phi_j(x_i)$, e.g. a 4-edges black vertex for $x_1$ with one 1-colored edge (solid line), two 2-colored edges (dotted lines) and a 3-colored edge (dashed line).
\item Running recursively from the left to the right of $\Gamma$, for each term inside brackets and bars (e.g. $\psi_1(x_2)\psi_1(x_1)\psi_1(x_2)$, then $\psi_2(x_1)\psi_1(x_2)\psi_2(x_3)\psi_1(x_2)\psi_2(x_1)$...), select randomly according to the colors and powers showing up in the monomials outgoing edges of the corresponding vertices and order them according to the order of the appearance of the colors in the (noncommutative !) monomial (e.g. select a 2-colored edge from the $x_2$ vertex, label it $a$; a 1-colored edge from the $x_1$ vertex, label it $b$; a 2-colored edge from the $x_2$ vertex, label it $c$). Connect these edges to a new white vertex (e.g. a new white vertex with 3 outgoing colored and ordered edges, where the order is defined by the labels).
\end{enumerate}

These edge-colored and locally ordered bipartite graphs will be called from now \it free interaction graphs\rm . 

The usual simplifications are possible in many cases of interest, we do not detail them and simply mention that, when the only monomials showing up in brackettings are of the form $\psi_i(x_m)\psi_i(x_l)$, the graphical rule amounts to consider ``classical'' Feynman graphs with labelled vertices and colored and directed propagators.

The noncommutative nonlocal case could be treated similarly by mixing the conventions for the noncommutative local and commutative local cases. The exercice is left to the reader.

\section{Symmetry factors and connectedness}{\label{symfac}}

This section is devoted to a technical but fundamental Lemma that connects the symmetry factors of graphs with the topological notion of connectedness. The lemma is the ground for the proof of linked cluster theorems and is particularly meaningful in the noncommutative case (Wightman fields).
We assume in this section that the Hopf algebra is an arbitrary free or free commutative combinatorial Hopf algebra which generators are primitive elements (this condition is satisfied by all the combinatorial Hopf algebras we have considered so far).

Let us introduce first a further notation. Let $x$ be a basis element (a commutative or noncommutative monomial in the generators) of a combinatorial Hopf algebra which support $S$ decomposes into a disjoint union $T\coprod V$ (the support is defined as for brackettings). We write $x_T$ and $x_V$ and call respectively $T$ and $V$-components of $x$ the two basis elements (possibly equal to zero) defined by $x_T\otimes x_V:= (P_T\otimes P_V)\circ \Delta (x)$, where $P_V$ stands for the projection on $H_V$ orthogonally to all the basis elements that do not belong to $H_V$. 

The reader can check that this definition amounts to the following: to get $x_T$, replace, in the expansion of $x$ as a monomial, all the $\phi_i(x_K), K\subset V $ by a 1 and all the $\phi_i(x_K), K\cap V\not=\emptyset, K\cap T\not=\emptyset $ by a zero (and similarly for $x_V$).

%Let us write $ev$ for the product map from $\PP^{\otimes n}$ or $\PP^{\otimes n}/S_n$ to $\PP$, so that e.g. for %$\Gamma=[\phi(x_1)^2\phi(x_2)|\phi(x_1)\phi(x_4)]$,
%$ev(\Gamma)=\phi(x_1)^3\phi(x_2)\phi(x_4)$.
%Let us write also $s_\Gamma$ for the symmetry factor of $\Gamma$ in $\GG(ev(\Gamma))$. 
%The following Lemma will prove extremely useful.

\begin{lem}\label{keylemma} Let $\Psi$ be a basis element and $ \Gamma=\Gamma_1\cdot \Gamma_2$ a bracketting or, equivalently, an interaction graph (of any type) such that $sup(\Gamma_1)\cap sup(\Gamma_2)=\emptyset$ and $s_{\Gamma}^\Psi\not=0$ (recall that $\cdot$ stands for the product of brackettings). Topologically, this amounts to assume that $\Gamma$ decomposes as a disjoint union of graphs.
Then:
\begin{itemize}
\item The basis elements $\Psi_{sup(\Gamma_1)}$ and $\Psi_{sup(\Gamma_2)}$ are non zero.
\item Moreover: $s_\Gamma^\Psi=s_{\Gamma_1}^{\Psi_{sup(\Gamma_1)}}s_{\Gamma_2}^{\Psi_{sup(\Gamma_2)}}$
\end{itemize}
In the commutative case, this last identity can be abbreviated to $s_\Gamma=s_{\Gamma_1}s_{\Gamma_2}$.
\end{lem} 

Indeed, since the $\phi_i(x_K)$ are primitive, the coproduct of a product, say $a_1...a_n$, of $\phi_i(x_K)$s is the sum of the tensor products $a_{i_1}...a_{i_k}\otimes a_{j_1}...a_{j_{n-k}}$, where $\{i_1,...,i_k\}$ is a (ordered) subset of $[n]$ and $\{j_1,...,j_{n-k}\}$ its (ordered) supplement. The hypothesis $s_{\Gamma}^\Psi\not=0$ ensures that $sup(\psi)= sup(\Gamma_1)\coprod sup(\Gamma_2)$, and that in $\psi$, there is no factor $\phi_i(x_K)$ with $ K\cap V\not=\emptyset$ and $ K\cap T\not=\emptyset $. The first assertion follows.

To prove the second identity, let us first notice that $s_\Gamma^\Psi=s_\Gamma^{\Psi_{sup(\Gamma_1)}\Psi_{sup(\Gamma_2)}}$. Indeed, let us use the same notation as previously and write $\Psi=a_1...a_n$. The coproduct and its iterations are constructed by extracting disjoint subsequences out of the ordered sequence of the $a_i$s. On the other hand, the basis elements showing up in $\Gamma_1$ and $\Gamma_2$ belong to disjoint sets -the relative ordering of the basis elements with support in $sup(\Gamma_1)$ and in $sup(\Gamma_2)$ in the expansion of $\Psi$ do therefore not matter, which proves the identity.

We can now assume without restriction (because of the cocommutativity) that $\Gamma=[S_1,...,S_k,T_1,...,T_l]$ with $\Gamma_1=[S_1,...,S_k]$ $\Gamma_2=[T_1,...,T_l]$.
Besides, since $\Delta$ is an algebra map, we have:
$$\Delta^{[k+l]}(\Psi_{sup(\Gamma_1)}\Psi_{sup(\Gamma_2)})=\Delta^{[k+l]}(\Psi_{sup(\Gamma_1)})\Delta^{[k+l]}(\Psi_{sup(\Gamma_2)}).$$
The multiplicity $\mu_\Gamma:=(k+l)!s_\Gamma^\Psi$ of $\Gamma$ in $\Delta^{[k+l]}(\Psi)$ is therefore obtained by summing the coefficients 
of the tensor products $(X_{\sigma(1)},...,X_{\sigma(k+l)})$ in $\Delta^{[k+l]}(\Psi)$, where $\sigma$ runs over $S_{k+l}/Stab((X_1,...,X_{k+l}))$ and $(X_1,...,X_{k+l})=(S_1,...,S_k,T_1,...,T_l)$ (here, $Stab((X_1,...,X_{k+l}))$ stands for the stabilizer of $(X_1,...,X_{k+l})$ in $S_{k+l}$). However, since the coproduct is cocommutative, these coefficients are all equal and one can restrict the computation to the tensor products $(S_{\beta(1)},...,S_{\beta(k)},T_{\alpha(1)},...,T_{\alpha(l)})$, where $\alpha$ and $\beta$ run over permutations in $S_k/Stab((S_1,...,S_k))$ and $S_l/Stab((T_1,...,T_l))$, and multiply the result by the number of $k$-element subsets in $[k+l]=\{1,...,k+l\}$.
We get finally:
$$\mu_\Gamma={{k+l}\choose{k}}\mu_{\Gamma_1}\mu_{\Gamma_2},$$
that is, $s_\Gamma^\Psi=s_{\Gamma_1}^{\Psi_{sup(\Gamma_1)}}s_{\Gamma_2}^{\Psi_{sup(\Gamma_2)}}$ or, in words (and slightly abusively), ``the symmetry factor of an interaction graph is the product of the symmetry factors of its connected components''. Notice that the property holds true also in nonlocal and/or noncommutative cases.

\section{Amplitudes and Feynman rules}\label{machin}

A linear form on a combinatorial Hopf algebra is \it unital \rm if $\rho(1)=1$ and \it infinitesimal \rm if $\rho(1)=0$.

Let us recall, for example, how linear forms on $\mathcal B_k^X$, $X=\{x_1,...,x_n,...\}$, are usually constructed. 
Let $T$ be an arbitrary finite sequence of integers. 
For any polynomial in $k$ variables, say $P(y_1,...,y_k)=\sum\limits_{1\leq i_1+...+i_k\leq n}p_{i_1,...,i_k}y_1^{i_1}...y_k^{i_k}$, we write $P(T)$ for the polynomial $\prod\limits_{t\in T}P(\phi_1(x_t),...,\phi_k(x_t))\in{\mathcal B}_k^X$. 
One can think of $P$ as the interacting part of a Lagrangian.
Natural forms should then be thought of as related physical amplitudes.
For example, for $k=1$, a typical $P$ is $\lambda\frac{y^4}{4!}$, corresponding to the $\phi^4$ theory (see \cite{KS} for details). 
The Green functions of this theory are computed via the
formula
\begin{eqnarray*}
G(x_1,\dots,x_n) := 
\frac{\rho\Big(\phi(x_1)...\phi(x_k)e^{i\int \phi^4(x) dx}\Big)}
{\rho\Big(e^{i\int \phi^4(x) dx}\Big)},
\end{eqnarray*}
where $\rho$ denotes the vacuum expectation value
of the time-ordered product of fields
(see section~\ref{QFTsect}).

Let us treat now a radically different example to show the ubiquity of the approach. Let here the role of $P$ be taken by $H_I(t)$, the interacting Hamiltonian of time-dependent perturbation theory (see e.g. our \cite{BP-09,BMP-10}): $H_I(t):=e^{iH_0t}Ve^{-iH_0t}e^{-\epsilon |t|}$.
Let $e_1,...,e_n,...$ be the eigenvectors of $H_0$ with eigenvalues $\lambda_1<\lambda_2\leq ...\leq \lambda_n\leq ...$. We assume for simplicity that the ground state is non degenerate ($\lambda_1\not=\lambda_2$), although the following reasoning holds in full generality due to \cite{BP-09,BMP-10}. The computation of the ground state of the perturbed Hamiltonian $H_0+V$ relies on the computation of the quantities such as:
$Y=<e_1|H_I(t_1)...H(t_p)|e_1>$, where $t\geq t_1>...>t_p> -\infty$. 
We set $k=\infty$ and $\phi_{2p}(t_j):=e^{-i\lambda_pt_j-\frac{\epsilon}2|t_j|}<e_p|$, $\phi_{2p+1}(t_j):=e^{+i\lambda_pt_j-\frac{\epsilon}2|t_j|}|e_p>$. Then, $H_I(t)=\sum_{i,j}V_{i,j}\phi_{2j+1}(t)\phi_{2i}(t)$, where $V_{i,j}:=<e_j|V|e_i>$. 
The unital form corresponding to the computation of $Y$ is given simply by the form on ${\mathcal F}_\infty^{]-\infty ,t]}$:
$$\rho(\phi_{i_0}(t_0)...\phi_{i_{2k+1}}(t_{2k+1})):=\prod\limits_{0\leq j\leq k}V_{i_{2j+1},i_{2j}}(\phi_{i_{2j}}(t_{2j})|\phi_{i_{2j+1}}(t_{2j+1})),$$
where $(\phi_{i_{2j}}(t_{2j})|\phi_{i_{2j+1}}(t_{2j+1})):=\phi_{i_{2j}}(t_{2j})\phi_{i_{2j+1}}(t_{2j+1})$ if $i_{2j}$ is even and $i_{2j+1}$ odd and zero else.
The value of the form $\rho$ on odd products is $0$. Of course, this example is purely didactical and for such a computation the use of the formalism developped in the present article is largely pointless. It becomes useful when the situation gets more involved. Actually, the simple requirement of taking efficiently into account the divergences arising from the adiabatic expansion may involve advanced combinatorial techniques, see, besides the articles already quoted, our \cite{BDPZ10}.

It is well-known that, in many situations, Green functions such as the ones of the $\phi^4$ theory split into components parametrized by Feynman diagrams. This property also holds for more complex theories and is best explained through Hopf algebraic computations.
Recall first that, since $H$ is a Hopf algebra, the set $ H^\ast$ of
linear forms on $H$ is equipped with an (associative; commutative if $H$
is cocommutative) ``convolution'' product: 
$$\forall \rho,\mu \in  H^\ast, \rho\ast\mu (x):=\rho(x_{(1)})\mu(x_{(2)}),$$
where we used the Sweedler notation $\Delta (x)=x_{(1)}\otimes x_{(2)}$. 
Notice that, if $\rho$ and $\mu$ are infinitesimal forms, $\rho\ast\mu (x):=\rho(x_{\{1\}})\mu(x_{\{2\}})$, where we use the notation $\overline\Delta (x)=x_{\{1\}}\otimes x_{\{2\}}$.
By standard graduation arguments, the convolution logarithm of a unital form $\rho$ is a well-defined infinitesimal form $\tau$ on $\mathcal P$. We extend such a $\tau$ to a linear form (still written $\tau$) on $\PP^{\otimes n}$ (resp. $\PP^{\otimes n}_{sym}$) by: $\tau (x_1\otimes ...\otimes x_n):=\tau(x_1)...\tau(x_n)$ or $\tau[x_1|...|x_n]:=\tau(x_1)...\tau(x_n)$.

Summing up, we get, for $X$ an arbitrary monomial (basis element) in $H$, and since $\tau$ is an infinitesimal form:
\begin{prop}[Feynman diagrams/rules expansion] For an arbitrary unital form on $H$, we have:
$$\rho(X)=\exp^{\ast \tau}(X)=\tau\circ\GG (X),$$
or:
$$\rho(X)=\sum\limits_\Gamma s_\Gamma^X\tau(\Gamma),$$
where $\Gamma$ runs over all the brackettings (or interaction graphs) in the image of $X$ by the graphication map. 
\end{prop}
The map $\tau$ acting on the $\Gamma$s is called a Feynman rule. Applying Lemma~\ref{keylemma}, with $H$ as in Sect.~\ref{symfac} we get immediately (with self-explanatory notations):

\begin{lem}\label{syfactrule}
Assume that $\Gamma =\Gamma_1\cdot \Gamma_2$, then:
$$s_\Gamma^X\tau(\Gamma)=s_{\Gamma_1}^{X_1}\tau(\Gamma_1)s_{\Gamma_2}^{X_2}\tau(\Gamma_2).$$
\end{lem}

\section{The combinatorial linked cluster theorem}

The combinatorial linked cluster theorem expands a linear form on a combinatorial Hopf algebra into ``connected'' parts closely related to the topological notion of connectedness. In this section, we show that this expansion is a very general phenomenon related to M\"obius inversion in the partition lattice. Notations and conventions are in Sect.~\ref{symfac}.

We first recall some general facts on the partition lattice and M\"obius inversion that are familiar in combinatorics but probably not well-known by practitioners of Feynman-type diagrammatics and linked cluster theorems (the particular case of M\"obius inversion for the partition lattice we are interested in here seems due to Sch\"utzenberger, we refer to \cite{Comtet} for further details and references on the subject).

For an arbitrary set $S$, partitions $t:=\{T_1,...,T_k\}$ of $S$ (that is: $T_1\coprod ...\coprod T_k=S$, where $\coprod$ stands for the disjoint union) are organized into a poset (partially ordered set, this poset is actually a lattice -two elements have a $max$ and a $min$, this follows easily from the definition of the order). We write $|t|$ for the length of the partition (so that $|t|=k$) and abbreviate the partitions of minimal and maximal length, respectively $\{S\}$ and $\{\{s\},\ s\in S\}$ to  $\hat 1$ and $\hat 0$.
The subsets $T_i$ are called the blocks, and the order is defined by
refinement: for any partitions $t$ and $u$, $t\leq u$ if and only if each block of ${t}$ is contained in a block of $u$.

The functions $f(x,y)$ on the partition lattice such that $f(x,y)\not=0$ only if $x\leq y$ form the \it incidence algebra \rm of the lattice. 
The (associative) product is defined by convolution: $(f\ast g)(x,y):=\sum\limits_{x\leq z\leq y}f(x,z)g(z,y)$.
The identity of the algebra is Kronecker delta function: $\delta(x,y):=1$ if $x=y$ and $:=0$ else.
The zeta function $\zeta(x,y)$ of the lattice
is defined to be equal to 1 if $x\leq y$ and 0 otherwise. 
The
M\"obius function $\mu(x,y)$ is defined to be the inverse of the zeta function for the convolution
product. It can be computed explicitly: for $x\leq t$, where $t=\{T_1,...,T_k\}$, we have:
$$\mu(x,t)=(-1)^{|x|+|t|}(n_1-1)!...(n_k-1)!,$$
where $n_i$ is the number of blocks of $x$ contained in $T_i$.
Using the identities $\mu\ast \zeta(\hat 0,\hat 1)=\zeta\ast\mu (\hat 0,\hat 1)=\delta (\hat 0,\hat 1)=0$, we recover in particular the useful combinatorial formulas:
$$\sum\limits_{0\leq k\leq |S|}\sum\limits_t(-1)^{|t|+|S|}(t_1-1)!...(t_k-1)!=0$$
where $t$ runs over the partitions of length $k$ of $S$ and $t_i$ stands for the number of elements in the $i$-th block $T_i$ of $t$ and, with the same conventions on $t$,
\begin{equation}\label{eqq}
\sum\limits_{0\leq k\leq |S|}\sum\limits_t(-1)^{|t|+1}(|t|-1)!=0
\end{equation}

The key application of these notions is to inclusion/exclusion computations in the partition lattice. Namely, for an arbitrary function $h(x)$ on the lattice, let us set: $h(y)=\sum\limits_{x\leq y}\widetilde{h}(x)$. This formula defines uniquely $\widetilde{h}$ and, in the convolution algebra:
$$(h=\widetilde{h}\ast \zeta )\Leftrightarrow (\widetilde{h}=h\ast \mu )$$
so that $\widetilde{h}(y)=\sum\limits_{x\leq y}h(x)\mu(x,y)$.

Let now $x$ be a basis element of $H$ with support written $S$, where $H$ is as in Sect.~\ref{symfac}. Let $\rho$ be a unital form on $H$. Recall the notation $x_T$ denoting the ``T-component'' of $x$ for an arbitrary $T\subset S$. For an arbitrary set partition $t=\{T_1,...,T_k\}$ of $S$, we extend $\rho$ to a function $\rho^x$ on partitions of $S$ and set:
$$\rho^x(t):=\rho(x_{T_1})...\rho(x_{T_k}).$$
Recall also the decomposition $\rho(x)=\sum_\Gamma s_\Gamma^x\tau(\Gamma)$. The $\Gamma$s are represented by diagrams, among which some are connected. We set $\rho_{conn}(x):=\sum_{\Gamma_c} s_{\Gamma_c}^x\tau(\Gamma_c)$, where the $\Gamma_c$ run over \it connected \rm diagrams.
We can then apply the machinery of inclusion/exclusion to $\rho^x$ and define $\widetilde\rho^x$. The combinatorial linked cluster theorem relates $\rho_{conn}$ and $\widetilde\rho^x$:

\begin{thm}[Combinatorial linked cluster theorem]
We have, for an arbitrary basis element in a free or free commutative combinatorial Hopf algebra which generators are primitive elements:
$$\rho_{conn}(x)=\widetilde\rho^x(\hat 1).$$
\end{thm}

We have indeed:
$$\widetilde\rho^x(\hat 1)=\sum\limits_{t}\rho^x(t)\mu(t,\hat 1)$$
where $t$ runs over the partitions of $S$ and (with the usual notations) $\mu(t,\hat 1)=(-1)^{|t|+1}(|t|-1)!$.
On the other hand, $\rho^x(t)=\rho(x_{T_1})...\rho(x_{T_k})$ and $\rho(x_{T_i})=\sum\limits_{\Gamma_i}s_{\Gamma_i}^{x_{T_i}}\tau(\Gamma_i)$. Lemma~\ref{syfactrule} ensures that $s_{\Gamma_1\cdot ...\cdot\Gamma_k}^x=s_{\Gamma_1}^{x_{T_1}}...s_{\Gamma_k}^{x_{T_k}}$, and we get finally:
$$\widetilde\rho^x(\hat 1)=\sum\limits_{T_1\coprod ...\coprod T_k=S}(-1)^{k+1}(k-1)!\sum\limits_{\Gamma_1,...,\Gamma_k}s_{\Gamma_1\cdot ...\cdot\Gamma_k}^x \tau(\Gamma_1)\cdot ...\cdot \tau(\Gamma_k),$$
where $\Gamma_i$ is a graph showing up in the expansion of $\rho(x_{T_1})$.

Let now $\Psi$ be an arbitrary graph with $s_\Psi^x\not=0$. The graph decomposes uniquely as a union of topologically disjoint graphs $\Psi_1\coprod...\coprod \Psi_n$, where $n$ is the number of connected components of $\Psi$. We write $S_i$ for $sup(\Psi_i)$ and $S_\Psi=\{S_1,...,S_n\}$.
We have to show that the coefficient of $\tau(\Psi)$ in the right hand side of the previous equation is equal to $s_\Gamma^x$ if $n=1$ and to 0 else.
The first property is immediate, since if $\Gamma$ is connected it appears only in the term associated to the trivial partition $\hat 1$ of $S$ and therefore with the coefficient $(-1)^20!s_\Gamma^x=s_\Gamma^x$. 

The second property is slightly less immediate, but follows from the
principles of M\"obius inversion together with Lemma~\ref{syfactrule}.
Notice first that $\Gamma$ appears in the right hand side of the
equation in association to all partitions of which $S_\Psi$ is a refinement.
The coefficient of $\Gamma$ is therefore:
$$[\sum\limits_{s_\Psi\leq t\leq \hat 1}(-1)^{|t|+1}(|t|-1)!]s_\Gamma^x$$
which is zero as a consequence of the identity $\zeta\ast\mu (S_\Psi,\hat 1)=\delta(S_\Psi,\hat 1)=0$ in the partition lattice.

\section{The functional linked cluster theorem}
A linear form $\rho$ on a free or free commutative combinatorial Hopf algebra $H$ is called \it symmetric \rm if it is invariant by a bijective relabelling of the variables $x_i$ -so that $\rho(\phi_1(x_2)^8\phi_3(x_5)^2\phi_2(x_9)^3)=\rho(\phi_1(x_4)^8\phi_3(x_2)^2\phi_2(x_1)^3)$, and so on. 
When $X$ is ordered, the form is called \it quasi-symmetric \rm if it is invariant by a (strictly) increasing relabelling of the variables, so that e.g. $\rho(\phi_1(x_2)^8\phi_3(x_5)^2\phi_2(x_9)^3)=\rho(\phi_1(x_4)^8\phi_3(x_5)^2\phi_2(x_8)^3)$, but is not necessarily equal to $\rho(\phi_1(x_4)^8\phi_3(x_2)^2\phi_2(x_1)^3)$.

Let us consider now a symmetric or quasi-symmetric unital form $\rho$ on $H$, where $H$ is as in Sect.~\ref{symfac} with an \it infinite \rm ordered index set $X$. For notational simplicity, we will assume that $X=\NN$. That is,
$H$ is an algebra of polynomials (resp. of tensors) over a doubly indexed set of formal, commuting (resp. noncommuting), and primitive variables $\phi_{i}(x_S)$ where $S$ runs over finite subsets of $X$ and $i=1...n_k$, where $k=|S|$ and where the sequence of the $n_k$, $k\in \NN$ is a fixed sequence of integers.

We first generalize the construction of the ``interaction term'' $P$ in Sect.~\ref{machin} as follows. We let ${\mathcal P}= P(T)_{ T\subset X}$ be a family of elements of $H$ such that $P(T)$ is a polynomial (resp. a tensor) in the $\phi_{i}(x_S),\ S\subset T$.

\begin{dfn} We say that ${\mathcal P}$ is \it admissible \rm if and only if 

\begin{enumerate}
\item For any order-preserving bijection $\phi$ from $T$ to $R$, $\phi(P(T))=P(R)$.
\item For any $T$ and any partition $U\coprod V=T$ (where $U$ and $V$ inherit the natural order on $T$), we have:
$$\sum_b\mu_b(P_b)_{U}\otimes (P_b)_{V}=P(U)\otimes P(V),$$
where $P(T)=\sum_b\mu_bP_b$ is the unique decomposition of $P(T)$ as a linear combination of basis elements.
\end{enumerate}
\end{dfn}

We let the reader check that the map $P$ constructed in Sect.~\ref{machin} satisfies this requirement. 

The composition of $\rho$ with $P$ is a ``scalar species'': the value $\hat\rho (S):=\rho\circ P(S)$ depends only on the number of elements in $S$ (in the quasi-symmetric case, an increasing bijection induces the identity $\hat\rho(S)=\hat\rho(T)$) so that if we set $\hat\rho (|S|):=\frac{\hat\rho (S)}{|S|!}$, the scalar species $\hat\rho$ is entirely characterized by the formal power series 
$$\hat\rho (x):=\sum\limits_n\hat\rho(n)x^n.$$

This straightforward remark connects QFT and many-body theory with the various algebraic structures existing on the algebra of formal power series.
Although apparently uselessly pedantical, it is actually useful to understand how these structures connect to the ones existing on scalar species.

There exists a Hopf-like structure on linear combinations of finite sets (see e.g. \cite{PatrasSchocker,PatrasSchocker2} for various developments of these ideas and the related notion of twisted algebras). The coproduct is defined by:
$$\delta(S):=\sum\limits_{U\coprod T=S}U\otimes T$$
where $\coprod$ stands for the disjoint union,
whereas the product is simply induced by the disjoint union of sets (the product of two overlapping sets is not defined).
These maps induce a convolution product written $\odot$ (to distinguish it from the convolution product of forms on $H$) on scalar species: for $\alpha,\beta$ two scalar species we get:
$$\alpha\odot\beta (S):=\sum\limits_{U\coprod T=S}\alpha(U)\beta(T),$$
or: ${\alpha\odot\beta} (x)=\alpha(x)\beta(x)$.

\begin{thm}[General functional linked cluster theorem]
We have, for any unital natural form $\rho$ on $H$ and admissible $\mathcal P$:
$$\log(\hat\rho (x))=\sum\limits_n\sum\limits_{\Gamma_n^c}\frac{s_{\Gamma_n^c}^{n}}{n!}\tau(\Gamma_n^c)x^n,$$
where $\tau:=\log^\ast(\rho)$, $\Gamma_n^c$ runs over the connected Feynman diagrams with vertex set $[n]$ and $s_{\Gamma_n^c}^{n}:=\sum\limits_b\lambda_bs_{\Gamma_n^c}^{b}$, where $P([n])=\sum\limits_b\lambda_b b$ is the decomposition of $P([n])$ as a linear combination of basis elements.
\end{thm}

Proof. 
We have:
$$\log(\hat\rho (x))=(\log^\odot\hat\rho)(x)=\sum\limits_n\frac{(\log^{\odot}\hat\rho)([n])}{n!}x^n$$
$$=\sum\limits_n\frac{x^n}{n!}\sum\limits_{I_1\coprod ...\coprod I_k=[n]}\frac{(-1)^{k+1}}{k}\hat\rho (I_1)...\hat\rho (I_k)$$
$$=\sum\limits_n\frac{1}{n!}\sum\limits_k\frac{(-1)^{k+1}}{k}\sum\limits_{I_1\coprod ...\coprod I_k=[n]}\sum\limits_{\Gamma_{I_1},...,\Gamma_{I_k}}s_{\Gamma_{I_1}}^{i_1}...s_{\Gamma_{I_k}}^{i_k}\tau (\Gamma_{I_1})...\tau (\Gamma_{I_k})$$
$$=\sum\limits_n\frac{1}{n!}\sum\limits_k\frac{(-1)^{k+1}}{k}\sum\limits_{I_1\coprod ...\coprod I_k=[n]}\sum\limits_{\Gamma_{I_1},...,\Gamma_{I_k}}s_{\Gamma_{I_1}}^{i_1}...s_{\Gamma_{I_k}}^{i_k}\tau (\Gamma_{I_1}... \Gamma_{I_k})$$
where $\Gamma_{I_i}$ runs over the Feynman brackets in the expansion of $\hat\rho (I_i)$, $i_1:=|I_i|$ and $\Gamma_i\Gamma_j$ denotes the concatenation of two brackets (so that e.g. $[\phi(x_1)|\phi(x_5)^2\phi(x_8)][\phi(x_2)^3|\phi(x_1)]=[\phi(x_1)|\phi(x_5)^2\phi(x_8)|\phi(x_2)^3|\phi(x_1)]$).

Now, let $\Gamma=\Gamma_1\coprod ...\coprod \Gamma_p$ be the (unique) decomposition of a Feynman diagram showing up in the expansion of $\hat\rho ([n])$ into a product of connected (non empty) diagrams. According to Lemma~\ref{keylemma} and since $\mathcal P$ is admissible, for any partition $A_1\coprod ...\coprod A_l$ of $[p]$, we have $s_{\Gamma_{A_1}}^{a_1}...s_{\Gamma_{A_l}}^{a_n}=s_{\Gamma}^n$, where $\Gamma_{A_i}:=\coprod\limits_{j\in A_i}\Gamma_j$ and $a_i$ is the number of vertices of $\Gamma_{A_i}$. 

The Theorem amounts then to the following properties: the coefficient of $\tau(\Gamma)$ in $\log(\hat\rho (x))$ is $\frac{s_\Gamma^n}{n!}x^n$ if $p=1$ (that is if the graph is connected) and zero else. 
The first property (the connected case) is obvious from the expansion. Let us assume therefore that $p>1$. The property follows once again from the general properties of the partition posets: the equation~(\ref{eqq}) concludes the proof.

\section{Examples}
\subsection{Quantum Field Theory}
\label{QFTsect}
In the quantum theory of the scalar field, the underlying Hopf algebra is the bosonic algebra ${\mathcal B}_k^{X}$, where $k$ is the number of fields and the elements of $X$ stand for dummy position or momentum variables. 

A typical example is the $\phi^4$ (scalar) theory with $k=1$ (we write simply $\phi$ for $\phi_1$).
The form $\rho$ computes expectation values of time ordered products of
free fields over the vacuum:
$$\rho(\phi(x_1)...\phi(x_k)):=<0|T(\phi(x_1)...\phi(x_k))|0>.$$
Problems arise when some of the $x_i$s coincide; these problems are the subject of the renormalization theory, we do not address them here.
The physically interesting quantities are the interacting Green functions 
\begin{eqnarray}
G(x_1,\dots,x_n) := 
\frac{\rho\Big(\phi(x_1)...\phi(x_k)e^{i\int \phi^4(x) dx}\Big)}
{\rho\Big(e^{i\int \phi^4(x) dx}\Big)}.
\label{Greendef}
\end{eqnarray}

The key point is that $\rho=\exp^*\tau$, where $\tau$ is
zero if its argument has degree different from two and
$\tau\big(\phi(x)\phi(y)\big):=<0|T(\phi(x)\phi(y))|0>$
is the Feynman propagator. The convolution logarithm can 
then be written as a sum of Feynman diagrams, where the lines
represent Feynman propagators and the vertices represent
spacetime points $x_i$. It can be checked that the
standard Feynman rules of quantum field theory~\cite{Itzykson} are
exactly recovered by the convolution exponential~\cite{BrouderMN}.
The linked-cluster expansion provides a simple way to
deal with the denominator of eq.~(\ref{Greendef})~\cite{Gross}.

\subsection{Cumulants}

Let  $X_1,...,X_n,...$ be a sequence of random variables.
The underlying Hopf algebra for this example is once again the bosonic algebra ${\mathcal B}_1^{\NN}$. The form $\rho$ is defined by:
$$\rho (\phi(x_{i_1})...\phi(x_{i_k})):=E[X_{i_1}...X_{i_k}].$$
This example enters the general commutative local case. 

When all the $x_i$s are distinct, $\rho (\phi(x_{i_1})...\phi(x_{i_k}))$ can be expanded as a sum parametrized by Feynman graphs which are disjoint unions of elementary graphs made of distinct black vertices, each joigned to a unique white vertex. The connected Feynman graphs appearing in the expansion correspond to the cumulants $E_c[X_{i_1}...X_{i_k}]$.
The combinatorial linked cluster actually shows that this graphical expansion is equivalent to
the classical identity:
$$E[X_1...X_n]=E_c[X_1...X_n]+\sum\limits_{A_1\cup ...\cup A_k}\prod_{i=1}^kE_c[X_{a_1^i}...X_{a_{j_i}^i}],$$
where $A_1\cup ...\cup A_k$ runs over the proper partitions of $[n]$ and $A_i=\{a_1^i,...,a_{j_i}^i\}$.

When the $X_i$ are copies of a given random variable $X$
and setting $P([n]):=E[X_1...X_n]$ (which is admissible), we recover, using the functional linked cluster theorem:
$$<e^X>_c-1=\log (E(e^X)),$$
with the convention $<1>_c=1$ and $<X^n>_c:=E_c[X_1...X_n]$.

\subsection{Quantum field theory with initial correlation}
In solid state physics and quantum chemistry, the initial
state is generally different from the vacuum. The physically
relevant form becomes (with the same notation as in
the first example)
$$\rho(\phi(x_1)...\phi(x_k)):=<\Phi|T(\phi(x_1)...\phi(x_k))|\Phi>,$$
where $|\Phi>$ is a general state. It is also possible to
consider a mixed state instead of the pure state $|\Phi\rangle$.
Except for very special cases (quasi-free states for bosonic
fields and Slater determinants for fermionic fields)
the convolution logarithm $\tau$ of $\rho$ is then
more complicated than in the first example. In particular,
$\tau$ can be nonzero if its argument have degree different
from two. In quantum optics, expansions in terms of $\tau$ are known
as cluster expansions and they lead to much
better convergence properties~\cite{Kira-08}. 
For the fermionic fields, the convolution logarithms $\tau$ are equivalent to
the cumulants of the reduced density matrices, that
are strongly advocated by Kutzelnigg and
Mukherjee~\cite{Kutzelnigg,KutzMukh,Kutzelnigg-00,Kong-10}.

The diagrammatic expansions can then not be done anymore using Feynman diagrams constructed out of Feynman propagators: see e.g. our \cite{BP09} and require the full apparatus of generalized Feynman diagrams for commutative local case. 

\subsection{Non-Gaussian measures}

Perturbative expansions in statistical physics for measures of Gaussian type can be performed using the usual Feynman graphs of Sect. \ref{QFTsect}.
This is because the Wick theorem applies.
When dealing with arbitrary functional measures this is not the case any more: higher cumulants (i.e. higher truncated moments or truncated Schwinger functions) have to be taken into account.

Feynman graphs and linked-cluster theorems have been developed by Djah and coll.~\cite{Djah} in this framework.
They were extensively used in several problems of probability 
theory~\cite{Djah2,Gottschalk-07,Gottschalk-08}.
These Feynman diagrams are equivalent to those of a quantum field theory with 
initial correlation.

\subsection{Free probabilities}

Free probabilities deal with the noncommutative local case and study linear forms on the tensor algebra ${\mathcal F}_1^\NN$. In general, the  graphs required to study such forms are free interaction graphs.

In practice, the theory of free probabilities focus often on linear forms with particular properties. This allows for various simplifications and typical properties as far as the corresponding cumulant expansions and their diagrammatic expansions are concerned. In particular, the Speicher's notion of free (or noncrossing) cumulant is obtained from the moment generating function by M\"obius inversion with respect to the lattice of noncrossing partitions (and not with respect to the lattice of partitions), see e.g. \cite{lehner} for further details and references on the subject.

\subsection{Truncated Wightman distributions}
In axiomatic quantum field theory, the 
form used in section~\ref{QFTsect} are replaced by 
$$W(x_1,...,x_n):=<\Phi|\phi_H(x_1)...\phi_H(x_k)|\Phi>,$$
where the operator product is used instead of the
time-ordered product, the fields are written in the
Heisenberg picture and $|\Phi\rangle$ is the ground
state of the interacting system.
Such functions are called Wightman distributions or 
correlation functions.
The main difference with the standard case is that
the Wightman distributions are not symmetric, the
order of the arguments is fixed.
Still, there is a perturbation theory of Wightman functions
that leads to non-commutative Feynman diagrams~\cite{Ostendorf-84}.
Their combinatorics is the same as for the standard case~\cite{BFK}.
Thus, the corresponding convolution logarithm $\tau$
is again zero if its argument is not of degree two,
but now 
$\tau\big(\phi(x),\phi(y)\big)
=\langle 0|\phi(x)\phi(y)|0\rangle$ is different
from $\tau\big(\phi(y),\phi(x)\big)$.

However, it is also possible to work at the non-perturbative
level and to define the form
$\rho\big(\phi(x_1)\dots\phi(x_n)\big):=W(x_1,\dots,x_n)$.
In that case, the convolution logarithm $\tau$ is
generally not zero if its argument is not of degree two
and $\tau\big(\phi(x_1)\dots\phi(x_n)\big)$
is now called a truncated Wightman distribution.

The definition of truncated Wightman distributions
was first given by Rudolf Haag in 1958~\cite{Haag58}.
We follow (up to the order of the variables) the definition
of Sandars' paper~\cite{Sanders-10}: ``For $n\ge1$ we let $\mathcal{P}_n$ denote
the set of all partitions of the set $\{1,\dots,n\}$ into pairwise disjoint
subsets, which are ordered from low to high. If $r$ is an ordered set in the
partition $P\in \mathcal{P}_n$ we write $r\in P$ and we denote the elements
of $r$ by $r(1)<\dots <r(|r|)$, where $|r|$ is the number of elements
of $r$. The truncated $n$-point distributions $\omega_n^T$,
$n\ge1$ of a state $\omega$ are defined implicitly in terms
of the $n$-point distributions 
\begin{eqnarray*}
\omega\big(\varphi(x_1),\dots,\varphi(x_n)\big)
&=& \sum_{P\in\mathcal{P}_n} \prod_{r\in P}
  \omega^T_{|r|}(x_{r(1)},\dots,x_{r(|r|)}).
\end{eqnarray*}
This is exactly the relation between $\rho=\exp^*\tau$ and 
$\tau$ for non-commuting variables.

From the physical point of view, the truncation procedure 
eliminates the contribution 
of the vacuum state as an intermediate state~\cite[p.~271]{Epstein}.
Truncated distributions have many desirable properties.
For instance, they decrease much faster than 
Wightman distributions at large space-like separation~\cite{Araki-60}.

\subsection{Nonrelativistic systems with Coulomb interaction}

Let us neglect here the problem of dealing with the Fermi statistics (which amounts essentially to introducing the correct signs in the definition of the Hopf algebra structures, see e.g. \cite{Cassam,BrouderQG}). 

Let us consider $n$ electrons in a quantum system
of non-relativistic electrons with a Coulomb interaction
in the external field generated by nuclei.
This is the standard approach of quantum chemistry 
and solid-state physics.

We assume that the non-interacting state can be 
described by a Slater determinant and that the
particle-hole transformation was used to
deal with occupied states. 
The form is defined as in section~\ref{QFTsect}
and the Green functions are now
$$
G(x_1,\dots,x_{n},y_1,\dots,y_n) := 
\frac{\rho\Big(\psi(x_1)\dots\psi(x_n)\psi^\dagger(y_1)
\dots\psi^\dagger(y_n)e^{-i I}\Big)}
{\rho(e^{-i I})},$$ 
where
$$
I=e^2
\int d t d\mathbf{r} d \mathbf{r'}
\frac{\psi^\dagger(\mathbf{r},t)\psi^\dagger(\mathbf{r}',t)
\psi(\mathbf{r}',t)\psi(\mathbf{r},t)}{8\pi\epsilon_0
  |\mathbf{r}-\mathbf{r}'|}.$$

The main difference with quantum field theory is that
the interaction is not local. Still, in the
linked-cluster expansion, we want to consider that
the points $\mathbf{r}$ and $\mathbf{r}'$ in $I$
are connected. In diagrammatic terms, 
$\mathbf{r}$ and $\mathbf{r}'$ are connected
by a wavy line. We introduced the tripartite
graphs to deal with this important case.

% BibTeX users please use
 \bibliographystyle{unsrt}
 \bibliography{triple}

\end{document}